\documentclass{INTERSPEECH2023}

\interspeechcameraready

\title{FOOCTTS: Generating Arabic Speech with Acoustic Environment for Football Commentator}
\name{Massa Baali$^1$, Ahmed Ali$^2$}
\address{
  $^1$Mohamed bin Zayed University of Artificial Intelligence, Abu Dhabi, UAE\\
  $^2$Qatar Computing Research Institute, HBKU, Doha, Qatar }
\email{massa.baali@mbzuai.ac.ae, amali@hbku.edu.qa}

\begin{document}

\maketitle
 
\begin{abstract}
This paper presents FOOCTTS, an automatic pipeline for a football commentator that generates speech with background crowd noise. The application gets the text from the user, applies text pre-processing such as vowelization, followed by the commentator's speech synthesizer. Our pipeline included Arabic automatic speech recognition for data labeling, CTC segmentation, transcription vowelization to match speech, and fine-tuning the TTS. 
Our system is capable of generating speech with its acoustic environment within limited 15 minutes of football commentator recording. Our prototype is generalizable and can be easily applied to different domains and languages.

\end{abstract}

\section{Introduction}
Generating speech with its corresponding acoustic environment remains a big challenge. Traditional text to speech (TTS) systems require high-quality, large-scale spoken-written pair training examples. However, the quality of TTS is still not sufficient for the generation of speech with its corresponding acoustic environment, especially for under-resourced languages or dialects, unlimited domains, or different speaking styles. In this paper, we present the FOOCTTS web application, which allows the user to input text and generate speech with its corresponding acoustic environment, which is a football crowd noise in our case. We propose a strategy of pre-training with source domain data followed by fine-tuning with target domain data. Our approach is fully unsupervised, where we collected videos from Youtube for a football commentator in the Tunisian dialect. Then, we labeled the data using a large pre-trained Arabic Automatic Speech Recognition (ASR). Our pipeline involved different preprocessing strategies such as CTC segmentation, voice activity detection (VAD),
and transcription vowelization to match the speech. In speech synthesis, we use transfer learning broadcast news domain and finetune it for the football commentating domain. Finally, we evaluate our model on a new text for one of the previous games and reproduced the commentating in a different style which was hard to tell that this was generated by a model.  Our main contributions fall in generating the acoustic environment with high-quality speech 
and with the data labeling using Arabic ASR. We use the following ESPNet-TTS recipe\footnote{ \url{https://github.com/espnet/espnet/tree/master/egs2/qasr_tts/tts1}}.   

\section{System Architecture}
\begin{figure}[t]
    \centering
\centerline{\includegraphics[width=9cm]{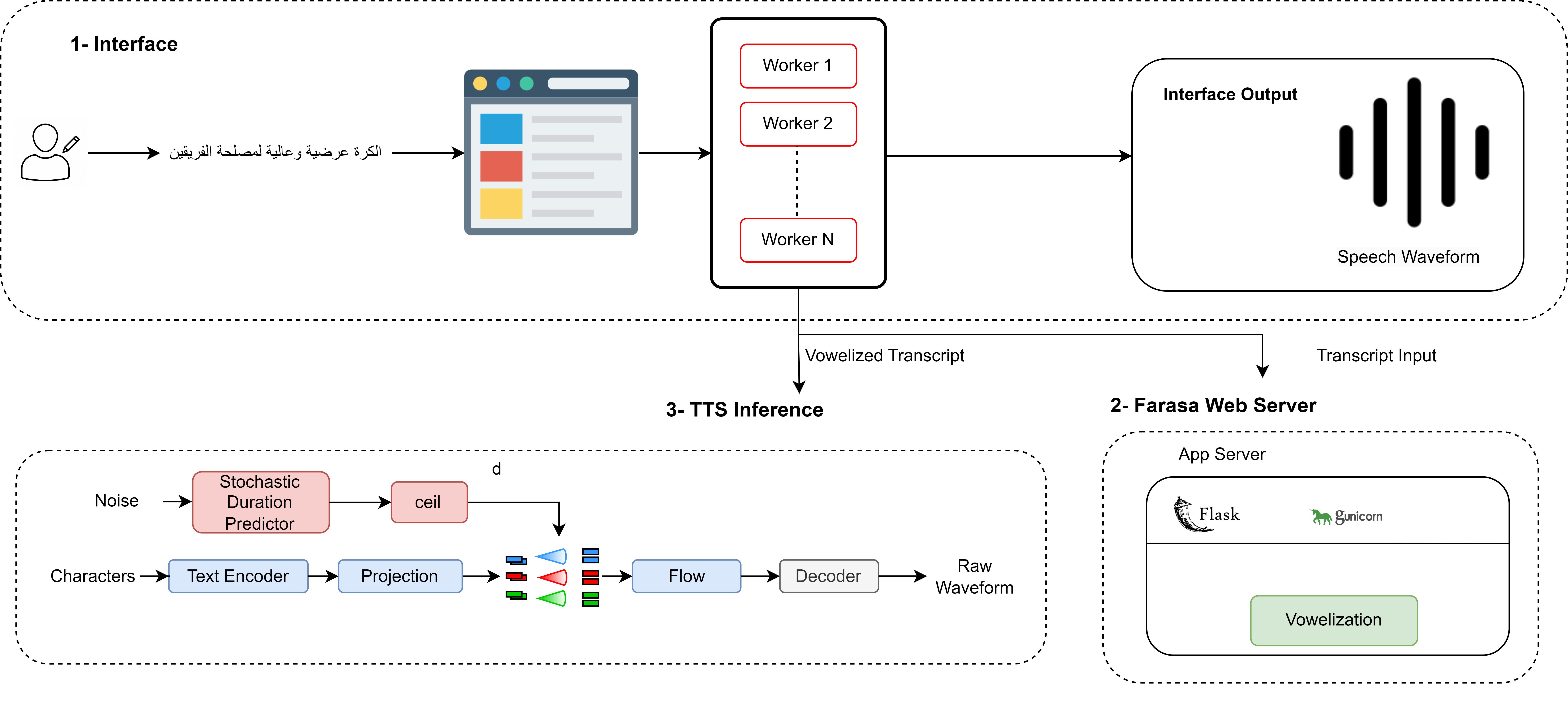}}
    \caption{System Overview}
    \label{fig:pipeline}
\end{figure}

The FOOCTTS system is composed of three independent components, namely: the web application, the TTS inference, and the Farasa \cite{abdelali2016farasa} toolkit application server. The complete system workflow diagram is shown in figure \ref{fig:pipeline}. It performs the following steps:
\begin{itemize}
  \item Capture input from user's keyboard through the interface.
  \item Spawn a worker that sends the text to the Farasa web server.
  \item Send vowelized text to TTS inference where we show in the figure the inference mode for our pre-trained variational inference with adversarial
learning for end-to-end text-to-speech (VITS) \cite{kim2021conditional} model.
  \item Play the final audio output on FOOCTTS.
\end{itemize}
\subsection{Web Application}
The web application is comprised of a front-end and a backend. The front-end presents the users with an interface to enter the text through a textbox. The front-end is primarily built with CSS and Javascript. For the backend, we used Flask and Nginx. A worker passes the text input to the Farasa server through its web API and receives back the vowelized transcript as an output. Once the transcribed output is received, the data is sent to the TTS inference through Flask API. 
\section{Implementation}
\begin{figure*}[t]
    \centering
\centerline{\includegraphics[width=17cm]{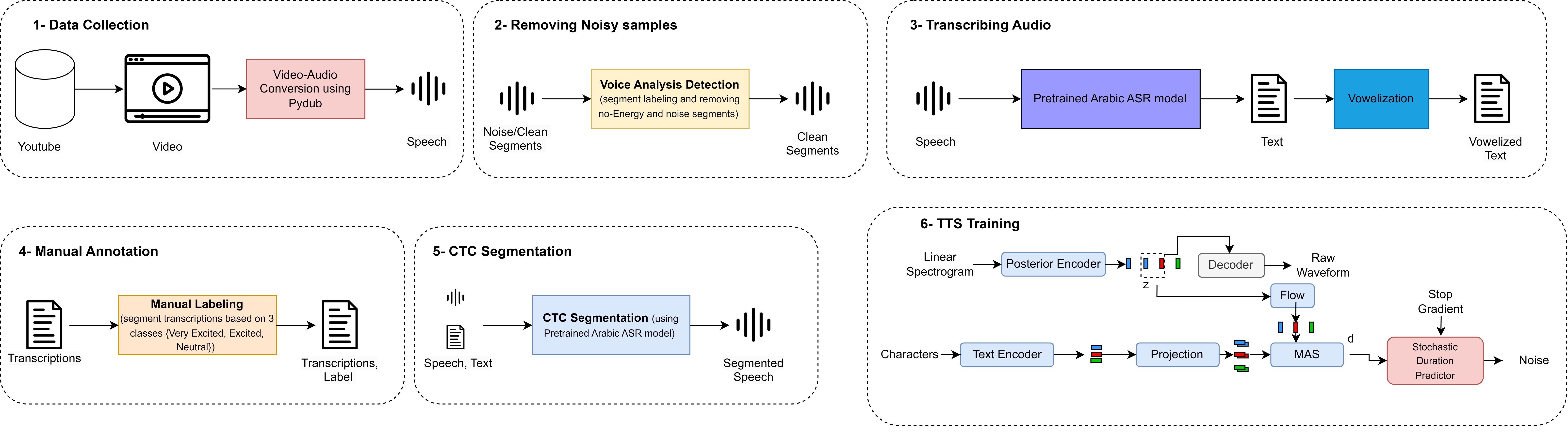}}
    \caption{Our proposed automatic pipeline}
    \label{fig:pipeline_train}
\end{figure*}

Figure \ref{fig:pipeline_train} illustrates the data collection, preprocessing strategies, and the overall architecture of the TTS model. 

\subsection{Building TTS Corpus from Football Commentator}
We collected videos from Youtube for football games for a Tunisian commentator reflecting the main use cases such as (shooting the ball, passing the ball, scoring, etc..). 

\subsubsection{Voice Activity  Detection}  \label{subsec:seg}
We extract the audio from the video samples at 22Khz. Then, we process the audio files of each football video. We investigate the impact of removing noise labels using voice activity detection (VAD). We use the inaSpeechSegmenter pre-trained model \cite{doukhan2018open} that extracts meta-data from each audio file; speech, noise, music and noEnergy (silence). We remove the noise, music, and noEnergy segments. 

\subsubsection{Transcribing Audio}
We use a large pre-trained code-switching Arabic ASR model to transcribe the speech samples. Our ASR model was able to transcribe the French words that are spoken by the commentator and the player's names. We replace the French words with Arabic words by applying transliteration using QCRI's transliteration API. \footnote{ \url{https://transliterate.qcri.org/}}

\subsubsection{Vowelization using Farasa}
In FOOCTTS, we employ the diacritizer provided by Farasa. One of the main issues is that vowelization is applied based on grammatical rules whereas if we want to perfect our output we should have the text vowelized according to how it was spoken. 

\subsubsection{Data Classification} \label{dataclass}
We apply manual annotation for 15 minutes of dialectal Arabic dataset. We ask the annotators to label the data into three different classes: 
neutral, excited, and very excited. We instruct them to classify the samples based on the pitch.  

\subsubsection{CTC Segmentation}
In the data classification part \ref{dataclass}, we got segmented transcriptions with their corresponding emotion. We investigate the impact of segmentation by applying CTC segmentation using the large pre-trained Arabic ASR model to segment the big audio chunks into smaller chunks to match them with the segmented transcriptions. Our goal is to have better training samples by having one sample labeled with one emotion in order to help the model learn faster. 

\subsection{Building Speech Synthesizer for Commentator} 
For baseline model, we start with previous work in \cite{baali2023unsupervised} and fine-tune it with 15 minutes of Tunisian commentator. 
\subsubsection{Model Architecture} \label{subsec: txt2wv}
We train VITS as our text to waveform (T2W) model which adopts a conditional variational autoencoder (CVAE) with normalizing flows and GAN-based optimizations. We train the model without the need for the two-stage pipelines which remain problematic because they require fine-tuning or training a neural vocoder. we show the architecture in this figure \ref{fig:pipeline_train}.
\subsubsection{TTS Training}
Our training process requires first training the text-to-waveform using 1 hour
of one anchor male speaker from QASR corpus \cite{mubarak2021qasr}. 
The text-2-wavform models were pre-trained as the character-based model with 1 hour from QASR dataset. We used 1150 utterances for the training and
25 utterances for the validation in pre-training. After pre-training, we fine-tune the model using commentator corpus. We used 25 utterances for development and 25 for testing and the rest for training.
\section{Conclusions}
This paper presents FOOCTTS, a speech synthesizer for a football commentator that generates speech with background crowd noise. Currently, the system works well without the user's ability to specify the emotion during inference time. For future work, we aim to allow the user to add the emotion with its corresponding text to generate speech in a certain tone. 
\bibliographystyle{IEEEtran}
\bibliography{mybib}

\end{document}